\documentclass[aps,prl,twocolumn,superscriptaddress,showpacs,floatfix]{revtex4}

\usepackage{graphicx}
\usepackage{dcolumn}
\usepackage{bm}
\usepackage{url} 
\usepackage{amsmath} 
\usepackage{amssymb} 

\begin{document}

\title{ \quad\\[1.0cm] Observation of the Doubly Cabibbo-Suppressed Decay $D^+_s\rightarrow K^+K^+\pi^-$}

\affiliation{Budker Institute of Nuclear Physics, Novosibirsk}
\affiliation{University of Cincinnati, Cincinnati, Ohio 45221}
\affiliation{T. Ko\'{s}ciuszko Cracow University of Technology, Krakow}
\affiliation{The Graduate University for Advanced Studies, Hayama}
\affiliation{Hanyang University, Seoul}
\affiliation{University of Hawaii, Honolulu, Hawaii 96822}
\affiliation{High Energy Accelerator Research Organization (KEK), Tsukuba}
\affiliation{Institute of High Energy Physics, Chinese Academy of Sciences, Beijing}
\affiliation{Institute of High Energy Physics, Vienna}
\affiliation{Institute of High Energy Physics, Protvino}
\affiliation{Institute for Theoretical and Experimental Physics, Moscow}
\affiliation{J. Stefan Institute, Ljubljana}
\affiliation{Kanagawa University, Yokohama}
\affiliation{Korea University, Seoul}
\affiliation{Kyungpook National University, Taegu}
\affiliation{\'Ecole Polytechnique F\'ed\'erale de Lausanne (EPFL), Lausanne}
\affiliation{Faculty of Mathematics and Physics, University of Ljubljana, Ljubljana}
\affiliation{University of Maribor, Maribor}
\affiliation{University of Melbourne, School of Physics, Victoria 3010}
\affiliation{Nagoya University, Nagoya}
\affiliation{Nara Women's University, Nara}
\affiliation{National Central University, Chung-li}
\affiliation{National United University, Miao Li}
\affiliation{Department of Physics, National Taiwan University, Taipei}
\affiliation{H. Niewodniczanski Institute of Nuclear Physics, Krakow}
\affiliation{Nippon Dental University, Niigata}
\affiliation{Niigata University, Niigata}
\affiliation{University of Nova Gorica, Nova Gorica}
\affiliation{Novosibirsk State University, Novosibirsk}
\affiliation{Osaka City University, Osaka}
\affiliation{University of Science and Technology of China, Hefei}
\affiliation{Seoul National University, Seoul}
\affiliation{Sungkyunkwan University, Suwon}
\affiliation{University of Sydney, Sydney, New South Wales}
\affiliation{Toho University, Funabashi}
\affiliation{Tohoku Gakuin University, Tagajo}
\affiliation{Department of Physics, University of Tokyo, Tokyo}
\affiliation{Tokyo Metropolitan University, Tokyo}
\affiliation{Tokyo University of Agriculture and Technology, Tokyo}
\affiliation{IPNAS, Virginia Polytechnic Institute and State University, Blacksburg, Virginia 24061}
\affiliation{Yonsei University, Seoul}
  \author{B.~R.~Ko}\affiliation{Korea University, Seoul} 
  \author{E.~Won}\affiliation{Korea University, Seoul} 
  \author{H.~Aihara}\affiliation{Department of Physics, University of Tokyo, Tokyo} 
  \author{K.~Arinstein}\affiliation{Budker Institute of Nuclear Physics, Novosibirsk}\affiliation{Novosibirsk State University, Novosibirsk} 
  \author{T.~Aushev}\affiliation{\'Ecole Polytechnique F\'ed\'erale de Lausanne (EPFL), Lausanne}\affiliation{Institute for Theoretical and Experimental Physics, Moscow} 
  \author{A.~M.~Bakich}\affiliation{University of Sydney, Sydney, New South Wales} 
  \author{V.~Balagura}\affiliation{Institute for Theoretical and Experimental Physics, Moscow} 
  \author{E.~Barberio}\affiliation{University of Melbourne, School of Physics,
    Victoria 3010} 
  \author{A.~Bondar}\affiliation{Budker Institute of Nuclear Physics,
    Novosibirsk}\affiliation{Novosibirsk State University, Novosibirsk} 
  \author{A.~Bozek}\affiliation{H. Niewodniczanski Institute of Nuclear Physics, Krakow} 
% ...
\collaboration{The Belle Collaboration}
\noaffiliation
 
\begin{abstract}
We report the first observation of the doubly Cabibbo-suppressed decay
$D^+_s\rightarrow K^+K^+\pi^-$ using 605 fb$^{-1}$ of data collected with the
Belle detector at the KEKB asymmetric-energy $e^+e^-$ collider. The branching
ratio with respect to its Cabibbo-favored counterpart
$\mathcal{B}(D^+_s\rightarrow K^+K^+\pi^-)$/$\mathcal{B}(D^+_s\rightarrow
K^+K^-\pi^+)$ is (0.229$\pm$0.028$\pm$0.012)\%, where the first uncertainty is
statistical and the second is systematic. We also report a significantly
improved measurement of the doubly Cabibbo-suppressed decay $D^+\rightarrow
K^+\pi^+\pi^-$, with a branching ratio $\mathcal{B}(D^+\rightarrow
K^+\pi^+\pi^-)$/$\mathcal{B}(D^+\rightarrow
K^-\pi^+\pi^+)$=(0.569$\pm$0.018$\pm$0.014)\%.
\end{abstract}


\maketitle

{\renewcommand{\thefootnote}{\fnsymbol{footnote}}}
\setcounter{footnote}{0}
Cabibbo-suppressed (CS) and doubly Cabibbo-suppressed (DCS) decays play an
important role in studies of charmed hadron dynamics. CS decays of nearly all
the charmed hadrons have been observed, while DCS decays have been observed for
only the $D^+$ and $D^0$ mesons. The na\"ive expectation for the DCS decay rate
is of the order of $\tan^4\theta_C$, where $\theta_C$ is the Cabibbo mixing
angle~\cite{CANGLE}, or about 0.29\%~\cite{LAMBDA} relative to its
Cabibbo-favored (CF) counterpart. Current measurements~\cite{PDG2008} roughly
support this expectation. It is natural to extend the searches for DCS decays
to other charmed hadrons in order to further understand the decay dynamics of
charmed hadrons and complete the picture. 

Furthermore, one expects that the branching ratio of $D^+\rightarrow
K^+\pi^+\pi^-$~\cite{CC} is about 2 $\tan^4\theta_C$ relative to its CF
counterpart since the phase space for $D^+\rightarrow K^-\pi^+\pi^+$ is
suppressed due to the two identical pions in the final state. This expectation
is consistent with current experimental results~\cite{PDG2008}. Therefore, we
also expect the branching ratio of $D^+_s\rightarrow K^+K^+\pi^-$ is about 1/2
$\tan^4\theta_C$ relative to its CF counterpart. Lipkin~\cite{LIPKIN} argues
that SU(3) flavor symmetry~\cite{SU3TRAN} implies
\begin{equation}
\frac{\mathcal{B}(D^+_s\rightarrow K^+K^+\pi^-)}{\mathcal{B}(D^+_s\rightarrow
  K^+K^-\pi^+)}\times\frac{\mathcal{B}(D^+\rightarrow
  K^+\pi^+\pi^-)}{\mathcal{B}(D^+\rightarrow K^-\pi^+\pi^+)}~=~\tan^8\theta_C,
  \label{EQ:TAN8}
\end{equation}
where differences in the phase space for CF and DCS decay modes cancel in the
ratios. The above relation does not take into account possible SU(3) breaking
effects that could arise due to resonant intermediate states in the three-body
final states considered here~\cite{LIPKIN}.

In this Letter, we report the first observation of the DCS decay
$D^+_s\rightarrow K^+K^+\pi^-$ and its inclusive branching ratio relative to
its CF counterpart, $D^+_s\rightarrow K^+K^-\pi^+$. We also report a new
measurement of the inclusive decay rate $D^+\rightarrow K^+\pi^+\pi^-$ relative
to its CF counterpart, $D^+\rightarrow K^-\pi^+\pi^+$. The current upper limit
on $\mathcal{B}(D^+_s\rightarrow K^+K^+\pi^-)/\mathcal{B}(D^+_s\rightarrow
K^+K^-\pi^+)$ is 0.78\% at the 90\% confidence level (C.L.)~\cite{FOCUS2} and
the world-average of the $D^+\rightarrow K^+\pi^+\pi^-$ branching ratio is
$\mathcal{B}(D^+\rightarrow K^+\pi^+\pi^-)/\mathcal{B}(D^+\rightarrow
K^-\pi^+\pi^+)$=(0.68$\pm$0.08)\%~\cite{PDG2008}. We also test the validity of
prediction~(\ref{EQ:TAN8}).

The data used in the analysis were recorded at the $\Upsilon(4S)$ resonance
with the Belle detector~\cite{BELLE} at the $e^+e^-$ asymmetric-energy collider
KEKB~\cite{KEKB}. It corresponds to an integrated luminosity of 605 fb$^{-1}$.

\begin{figure}[htbp]
\includegraphics[height=0.4\textwidth,width=0.5\textwidth]{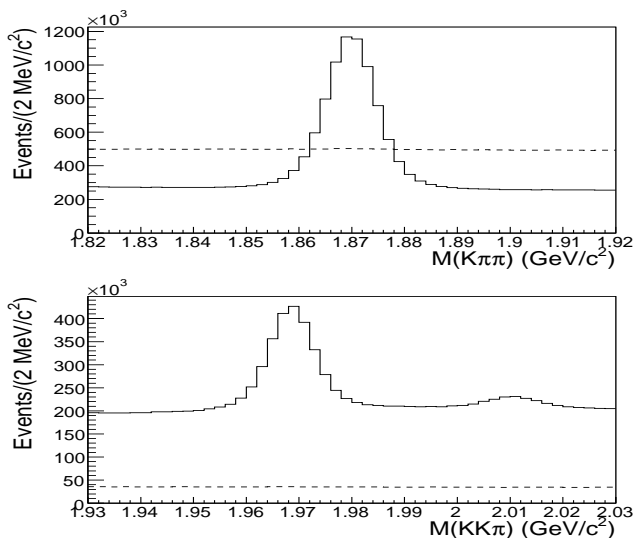}
\caption{$M(K\pi\pi)$ (top) and $M(KK\pi)$ (bottom) distributions after the
  initial selections. The solid curve shows the CF decay channel and the dashed
  shows the DCS decay channel in each plot.}
\label{FIG:M_STDCUT}
\end{figure}

$D^+$ and $D^+_s$ candidates are reconstructed using three charged tracks in
the event. The initial event selection is similar to that in other Belle
measurements. We require that the charged tracks originate from the vicinity of
the interaction point with impact parameters in the beam direction ($z$-axis)
and perpendicular to it of less than 4 cm and 2 cm, respectively. All charged
tracks are required to have at least two associated hits in the silicon vertex
detector~\cite{SVD2}, both in the $z$ and radial directions, to assure good
spatial resolution on the $D$ mesons' decay vertices. The decay vertex is
formed by fitting the three charged tracks to a common vertex and requiring a
C.L. greater than 0.1\%. Charged kaons and pions are identified requiring the
ratio of particle identification likelihoods,
$\mathcal{L}_K/(\mathcal{L}_K+\mathcal{L}_\pi)$, constructed using information
from the central drift chamber, time-of-flight counters, and aerogel Cherenkov
counters~\cite{PID}, to be larger or smaller than 0.6, respectively. In
addition, we require that the scaled momentum of the charmed meson candidate
$x_p=p^*/\sqrt{0.25\cdot E^2_{\rm CM}-M^2}$ be greater than 0.5 to suppress
combinatorial background as well as $D$ mesons produced in $B$ meson
decays. Here $p^*$ and $E_{\rm CM}$ are the charmed meson momentum and the
total $e^+e^-$ collision energy, calculated in the center-of-mass frame, and
$M$ is the reconstructed invariant mass of the
candidate. Figure~\ref{FIG:M_STDCUT} shows the $K\pi\pi$ and
$KK\pi$~\cite{CHARGED} invariant mass distributions after the initial
selections. The background levels are too high to observe DCS signals.

We then apply further selection criteria, which are optimized using real data
samples since there are some discrepancies between the Monte Carlo (MC)
simulation~\cite{BELLEMC} and the data in the relevant distributions. We use
10\% of the data sample for optimization and the remaining 90\% for the
measurement to avoid a possible bias when the same samples are used for both
optimization and the measurement. Hence the final selection criteria are
obtained in a blind manner. Assuming no signal in the DCS decay channel, we
maximize $\mathcal{N}_S/\sqrt{\mathcal{N}_B}$, where $\mathcal{N}_S$ is the CF
signal yield which has similar properties to the DCS signal and $\mathcal{N}_B$
is the background yield from the sideband regions in the DCS sample.

\begin{figure}[htbp]
\includegraphics[height=0.4\textwidth,width=0.5\textwidth]{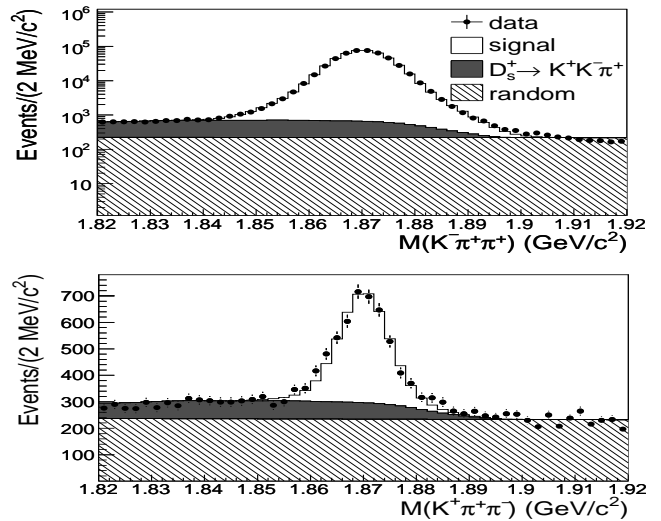}
\caption{Distributions of $M(K^-\pi^+\pi^+)$ (top) and $M(K^+\pi^+\pi^-)$
  (bottom). The $K^-\pi^+\pi^+$ distribution is shown on a semi-logarithmic
  scale to make the small background visible. Points with error bars show the
  data and histograms show the results of the fits described in the
  text. Signal, $D^+_s\rightarrow K^+K^-\pi^+$ background, and random
  combinatorial background components are also shown.}
\label{FIG:MKPIPI_FIN}
\end{figure}

\begin{figure}[htbp]
\includegraphics[height=0.4\textwidth,width=0.5\textwidth]{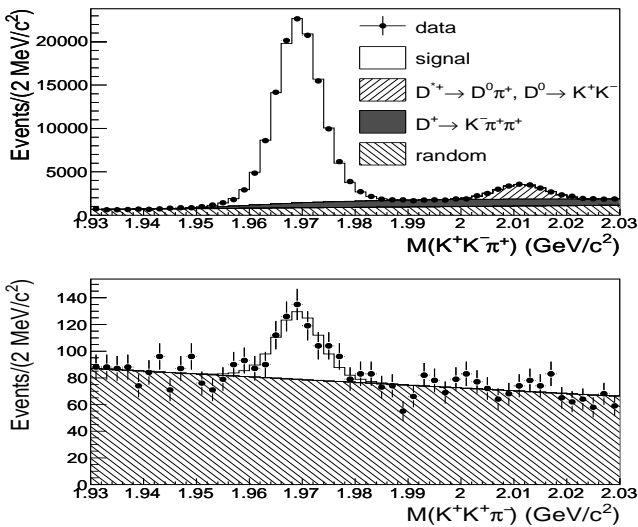}
\caption{Distributions of $M(K^+K^-\pi^+)$ (top) and $M(K^+K^+\pi^-)$
  (bottom). Points with error bars show the data and histograms show the
  results of the fits described in the text. Signal, $D^{*+}$ background
  ($D^{*+}\rightarrow D^0\pi^+$ where $D^0\rightarrow K^+K^-$), $D^+\rightarrow
  K^-\pi^+\pi^+$ background, and random combinatorial background components are
  also shown.}
\label{FIG:MKKPI_FIN}
\end{figure}

One of the selections related to the finite lifetime of charmed hadrons is the
reduced $\chi^2$ ($\chi^2$/d.o.f) for the hypothesis that the candidate tracks
for the charmed meson decay products arise from the primary vertex. The primary
vertex is obtained as the most probable point of intersection of the meson's
momentum vector and the $e^+e^-$ interaction region. Due to the finite lifetime
of $D^+$ and $D^+_s$ mesons their daughter tracks are not likely to be
compatible with the primary vertex. The second requirement uses the angle
between the charmed meson momentum vector, as reconstructed from the daughter
tracks, and the vector joining its production and decay vertices. In an ideal
case without resolution the two vectors would be parallel for the signal. The
reduced $\chi^2$ is required to be greater than 25 for $D^+$ and 5 for $D^+_s$
candidates and the angle is required to be less than 1$^{\circ}$ for $D^+$ and
2$^{\circ}$ for $D^+_s$ candidates. Tighter requirements on charged kaon
identification ($>$0.9) and $x_p$ ($>$0.7) are also chosen for the final
selection, which improves the signal sensitivity. After the additional and
tighter selection requirements described above, 9.57\% of $D^+$ and 10.71\% of
$D^+_s$ CF signal, and 0.06\% of $D^+$ and 0.24\% of $D^+_s$ DCS background
events are retained. In order to minimize systematic effects we choose the same
selection criteria for both DCS and CF decay channels. The $K\pi\pi$ and
$KK\pi$ invariant mass distributions after the final selections are shown in
Figs.~\ref{FIG:MKPIPI_FIN} and \ref{FIG:MKKPI_FIN} together with signal and
background parameterizations. A clear signal is observed in both DCS decay mass
distributions.

The $K\pi\pi$ and $KK\pi$ invariant mass distributions are fitted using the
binned maximum likelihood method. In all cases the signal probability density
function (PDF) is parameterized using two Gaussians with the same central
value. Due to $K/\pi$ misidentification the following reflection backgrounds
appear in the mass distributions. In $D^+\rightarrow
K^-\pi^+\pi^+$ (CF) and $K^+\pi^+\pi^-$ (DCS) decays there is a contribution
from misidentified $D^+_s\rightarrow K^+K^-\pi^+$ decay; in $D^+_s\rightarrow
K^+K^-\pi^+$ (CF) decay there is a contribution from misidentified
$D^+\rightarrow K^-\pi^+\pi^+$; and in $D^+_s\rightarrow K^+K^+\pi^-$ (DCS)
decay there is a contribution from misidentified $D^+\rightarrow K^+\pi^+\pi^-$
decay. The PDFs for the reflection backgrounds are determined from real data by
assigning the nominal pion (kaon) mass to a kaon (pion) track. The magnitude of
each of the reflection background contributions is a free parameter in the
fit. For the DCS $D^+_s$ channel, the $D^+\rightarrow K^+\pi^+\pi^-$
contribution is not incorporated in the fit since it is not significant, but
its effect is included as a systematic uncertainty due to fitting listed in
Table~\ref{TABLE:SYSTOTAL}. The $D^{*+}$ contribution ($D^{*+}\rightarrow
D^0\pi^+$ with $D^0\rightarrow K^+K^-$) in the CF $D^+_s$ channel is also
incorporated in the CF $D^+_s$ fit as an independent Gaussian component. A
linear function is used for the random combinatorial background for all
channels. All signal and background parameters for the CF channels are
floated. For the DCS channels the mass, width, and ratio of the two signal
Gaussians are fixed to the values obtained from the fits to distributions of CF
decays. Signal and background yields are left free in the fit. From the results
of the fits, shown in Figs.~\ref{FIG:MKPIPI_FIN} and \ref{FIG:MKKPI_FIN}, we
extract the signal yield for each channel, listed together with the
corresponding branching ratios in Table~\ref{TABLE:SUMMARY}.

The statistical significance of the $D^+_s\rightarrow K^+K^+\pi^-$ signal is
calculated using $-2\ln(\mathcal{L}_b/\mathcal{L}_{s+b})$ where $\mathcal{L}_b$
and $\mathcal{L}_{s+b}$ are the likelihood values of the fit, without and with
the signal PDF included, respectively. We find
$-2\ln(\mathcal{L}_b/\mathcal{L}_{s+b})$=83.2 with one degree of freedom used
to describe the DCS signal yield; we obtain a statistical significance
corresponding to 9.1 standard deviations.

In addition to the backgrounds mentioned above there is also the possibility of
double misidentification leading to contributions from CF events to the DCS
sample. MC simulation shows that such a contribution is flat in the invariant
mass distribution and is hence included in the combinatorial background
description.

The final states in this study have resonant substructure that can affect the
reconstruction efficiency. The resonances are relatively well known for the
decay modes other than $D^+_s\rightarrow K^+K^+\pi^-$. We used a coherent
mixture of resonant contributions according to~\cite{E691} to generate
$D^+\rightarrow K^-\pi^+\pi^+$ decays and calculate the reconstruction
efficiency. For the $D^+\rightarrow K^+\pi^+\pi^-$ and $D^+_s\rightarrow
K^+K^-\pi^+$ decays we used an incoherent mixture of intermediate
states~\cite{PDG2008}. Subsequently we varied the contributions of individual
intermediate states in a correlated manner, within the uncertainties of the
measured branching fractions. The efficiency calculated from the modified MC
sample differs from the original one by 1.5\% and 2.0\% for the $D^+\rightarrow
K^+\pi^+\pi^-$ and $D^+_s\rightarrow K^+K^-\pi^+$ decays, respectively, and the
difference was included in the systematic uncertainty of the
result. $D^+_s\rightarrow K^+K^+\pi^-$ decays were generated according to phase
space. For comparison, signal events were generated assuming either
$K^{*0}(892)K^+$ or $K^{*0}(1430)K^+$ intermediate states. The largest relative
difference in the efficiency (2.4\%) was included as a part of the systematic
uncertainty. Ratios of reconstruction efficiencies for DCS and CF decays are
found to be 1.042$\pm$0.008$\pm$0.016 and 0.963$\pm$0.010$\pm$0.030 for $D^+$
and $D^+_s$ decays, respectively, where the first uncertainty is due to the
finite MC simulation statistics and the second is the uncertainty in the
resonant structure of the final states.

With the efficiencies estimated above, we measure the inclusive branching
ratios of DCS decays relative to their CF counterparts summarized in
Table~\ref{TABLE:SUMMARY}. The product of the branching ratios for the two DCS
decay modes is found to be $\frac{\mathcal{B}(D^+_s\rightarrow
  K^+K^+\pi^-)}{\mathcal{B}(D^+_s\rightarrow
  K^+K^-\pi^+)}\times\frac{\mathcal{B}(D^+\rightarrow
  K^+\pi^+\pi^-)}{\mathcal{B}(D^+\rightarrow
  K^-\pi^+\pi^+)}~=~(1.57\pm0.21)\times\tan^8\theta_C$, where the error is the
total uncertainty.

\begin{table}[htbp]
\begin{center}
\caption{Measured branching ratios. $\mathcal{B}_{\rm rel.}$ is the branching
  ratio relative to $D^+\rightarrow K^-\pi^+\pi^+$ for the $D^+$ modes and
  $D^+_s\rightarrow K^+K^-\pi^+$ for the $D^+_s$ modes. The uncertainties in
  the branching ratios are statistical and systematic.}
\label{TABLE:SUMMARY}
\begin{tabular}{ccc} \hline \hline
Decay Mode           &$\mathcal{N}_{\rm signal}$&$\mathcal{B}_{\rm rel.}$(\%) \\ \hline
$D^+\rightarrow K^+\pi^+\pi^-$ &2637.7$\pm$84.4&0.569$\pm$0.018$\pm$0.014\\ 
$D^+\rightarrow K^-\pi^+\pi^+$ &482702$\pm$727 &100\\ \hline
$D^+_s\rightarrow K^+K^+\pi^-$ &281.4$\pm$33.8 &0.229$\pm$0.028$\pm$0.012\\ 
$D^+_s\rightarrow K^+K^-\pi^+$ &118127$\pm$452 &100\\ \hline \hline
\end{tabular}     
\end{center}
\end{table}

Several sources of systematic uncertainty cancel in the branching ratio
calculation due to the similar kinematics of CF and DCS decays (for example,
uncertainties in the tracking efficiencies and particle identification, since
the momenta of the final state tracks are almost identical). The stability of
the branching ratios against the variation of the selection criteria was
studied and we observed no changes greater than the expected statistical
fluctuations. The systematic uncertainties due to the variation of the fit
parameters are 1.9\% for $D^+$ and 4.2\% for $D^+_s$ branching ratios
measurements, respectively. Table~\ref{TABLE:SYSTOTAL} summarizes the
systematic uncertainties in the measurements of the branching ratios.
\begin{table}[htbp]
\begin{center}
\caption{Relative systematic uncertainties in percent, where
  $\sigma_{\mathcal{B}_{\rm rel.}(D^+)}$ and $\sigma_{\mathcal{B}_{\rm
  rel.}(D^+_s)}$ are systematic uncertainties for the branching ratio of $D^+$
  and $D^+_s$ DCS decays relative to their CF counterparts.}
\label{TABLE:SYSTOTAL}
\begin{tabular}{ccc} \hline \hline
Source                     &$\sigma_{\mathcal{B}_{\rm rel.}(D^+)}$ (\%) &$\sigma_{\mathcal{B}_{\rm rel.}(D^+_s)}$ (\%)\\ \hline
Fitting                    &1.9&4.2\\ 
MC Statistics              &0.8&1.0\\ 
Reconstruction Efficiency  &1.5&3.1\\ \hline
Total                      &2.5&5.3\\ \hline \hline
\end{tabular}     
\end{center}
\end{table}

Using the world-average values $\mathcal{B}(D^+\rightarrow
K^-\pi^+\pi^+)=(9.22\pm0.21)\%$ and $\mathcal{B}(D^+_s\rightarrow
K^+K^-\pi^+)=(5.50\pm0.28)\%$~\cite{PDG2008}, we obtain the absolute branching
fraction for each DCS decay channel. Table~\ref{TABLE:COMP_2008} shows the
results and the comparison to previous results.
\begin{table}[htbp]
\begin{center}
\caption{Absolute branching fraction for each decay mode and comparisons with
  previous measurements. The first uncertainties shown in the second column are
  the total uncertainties of our results and the second are the uncertainties
  in the average CF branching fractions used for normalization~\cite{PDG2008}.}
\label{TABLE:COMP_2008}
\begin{tabular}{ccc} \hline \hline
Branching Fraction           &Belle &World-Average~\cite{PDG2008} \\ \hline
$\mathcal{B}(D^+\rightarrow K^+\pi^+\pi^-)$&(5.2$\pm$0.2$\pm$0.1)$\times$10$^{-4}$&(6.2$\pm$0.7)$\times$10$^{-4}$ \\
$\mathcal{B}(D^+_s\rightarrow K^+K^+\pi^-)$&(1.3$\pm$0.2$\pm$0.1)$\times$10$^{-4}$&(2.9$\pm$1.1)$\times$10$^{-4}$ \\ \hline \hline
\end{tabular}     
\end{center}
\end{table}

To conclude, using 605 fb$^{-1}$ of data collected with the Belle detector at
the KEKB asymmetric-energy $e^+e^-$ collider we have observed for the first
time the decay $D^+_s\rightarrow K^+K^+\pi^-$ with a statistical significance of
9.1 standard deviations. This is the first DCS decay mode of the
$D^+_s$ meson. The branching ratio with respect to the CF decay is
(0.229$\pm$0.028$\pm$0.012)\%, where the first uncertainty is statistical and
the second is systematic. We have also determined the $D^+$ DCS decay branching
ratio, $\mathcal{B}(D^+\rightarrow K^+\pi^+\pi^-)$/$\mathcal{B}(D^+\rightarrow
K^-\pi^+\pi^+)$=(0.569$\pm$0.018$\pm$0.014)\%, where the first uncertainty is
statistical and the second is systematic, with a significantly improved
precision compared to the current world-average~\cite{PDG2008}. We find the
product of the two relative branching ratios,
$\frac{\mathcal{B}(D^+_s\rightarrow K^+K^+\pi^-)}{\mathcal{B}(D^+_s\rightarrow
  K^+K^-\pi^+)}\times\frac{\mathcal{B}(D^+\rightarrow
  K^+\pi^+\pi^-)}{\mathcal{B}(D^+\rightarrow K^-\pi^+\pi^+)}$ to be
(1.57$\pm$0.21)$\times\tan^8\theta_C$. This is consistent with SU(3) flavor
symmetry within three standard deviations; note that the effect of (different)
resonant intermediate states is not taken into account in the
prediction~\cite{LIPKIN}. An amplitude analysis on a larger data sample may
allow a more precise test of SU(3) flavor symmetry to be performed.

We thank the KEKB group for excellent operation of the accelerator, the KEK
cryogenics group for efficient solenoid operations, and the KEK computer group
and the NII for valuable computing and SINET3 network support. We acknowledge
support from MEXT, JSPS and Nagoya's TLPRC (Japan); ARC and DIISR (Australia);
NSFC (China); DST (India); MOEHRD and KOSEF (Korea); MNiSW (Poland); MES and
RFAAE (Russia); ARRS (Slovenia); SNSF (Switzerland);  NSC and MOE (Taiwan); and
DOE (USA).


\begin{thebibliography}{99}
\bibitem{CC}
Throughout this Letter the charge-conjugate state is implied unless stated
otherwise.
\end{thebibliography}
\end{document}